\UseRawInputEncoding 

\documentclass[twocolumn,showpacs,superscriptaddress,aps,pra]{revtex4-1}
\usepackage{epsfig}
\usepackage{subfigure}
\usepackage{amssymb}
\usepackage{graphicx}
\usepackage{color}
\usepackage{amsfonts}
\usepackage{amscd}
\usepackage{hyperref}
\usepackage{amsmath}
\usepackage{epstopdf}

\usepackage{amsmath}
\usepackage{graphicx}
\usepackage{txfonts}
\usepackage{textcomp}

\begin{document}

\title{Geuine tripartite entanglement in three-flavor neutrino oscillations}

\author{Yu-Wen Li}

\author{Li-Juan Li}

\author{Xue-Ke Song}
\email{songxk@ahu.edu.cn}

\author{Dong Wang}
\email{dwang@ahu.edu.cn}

\author{Liu Ye}

\affiliation{School of Physics and Optoelectronics Engineering, Anhui University, Hefei 230601, China}

\date{\today }

\begin{abstract}
The violation of Leggett-Garg inequalities tested the quantumness of neutrino oscillations (NOs) across macroscopic
distances. The quantumness can be quantified by using the tools of the quantum resource theories. Recently, a new genuine tripartite entanglement measure [S. B. Xie \emph{et al.}, Phys. Rev. Lett. \textbf{127}, 040403 (2021)], concurrence fill, is defined as the square root of the area of the concurrence triangle satisfying all genuine multipartite entanglement conditions. It has several advantages compared to other existing tripartite measures. Here, we focus on using concurrence fill to quantify the tripartite entanglement in three-flavor NOs.
Concurrence fill can reach its maximum $0.89$ for the experimentally-observed electron antineutrino oscillations, but it cannot for the muon antineutrino oscillations. In both cases,
we compare its performance with other three tripartite entanglement measures, including the generalized geometric measure (GGM), the three-$\pi$ entanglement, and the genuinely multipartite concurrence (GMC), in the neutrino propagation, and accordingly show that concurrence fill contains the most quantum resource.
Furthermore, concurrence fill and the three-$\pi$ entanglement are always smooth, while GGM and GMC measures have several sharp peaks. The genuine tripartite quantification
of the quantumness of three-flavor NOs represents the first step towards the further potential application of neutrinos on quantum information processing.

\end{abstract}

 \maketitle


\section{Introduction}
\label{sec1}

Neutrino is a Standard Model of neutral weakly interacting fermion \cite{Giunti}. It is the second most abundant particle in the Universe after photons of light.
Neutrino oscillation (NO) implies that the neutrino has a non-zero mass. In the framework of the simplest
standard model of three-neutrino mixing, three different flavors of neutrino are electron $e$, muon $\mu$, and tau $\tau$ leptons, in which the three flavor states
are unitary linear combinations of three mass eigenstates \cite{Camilleri:2008zz,Duan:2010bg}. NO shows that a given flavor may change into another flavor in the neutrino propagation.
The probability of measuring a particular flavor for a neutrino varies periodically as it propagates through space, and can be measured at the arbitrary time.
The values of the oscillations parameters have been measured and analyzed in both theory and experiment in recent years~\cite{abs15,PAda20,dayabaydata,Acero18,kdata21}.
Remarkably, oscillation probabilities of neutrino can be used to study the different properties from classical to quantum mechanical prediction of such an interesting
system.

As an analog of Bell's inequality in temporal interpretation, LGI study the correlations of a single system measured at different times, based on two assumptions
of macroscopic realism and non-invasive measurability \cite{Leggett,Emary,Budroni}.
It is shown that experimentally-observed neutrino oscillations can violate the classical limits imposed by the LGI \cite{Gangopadhyaytwo,Fu,Gangopadhyaythree,Naikoo19}, provide an evidence that quantum coherence can apply broadly to
microscopic systems over a  macroscopic distance. However, the violation of LGI can be taken as an effective
indicator for quantifying the amount of quantumness from the framework of quantum resource theories (QRTs). Recently, the quantification of quantumness
in terms of flavor oscillation probabilities of NOs
was investigated by quantities in QRTs,
including  entanglement \cite{mbf09,mbf14}, Svetlichny inequalities \cite{sba15,dixit}, entropic uncertainty relation \cite{dwfx,ljlf}, quantum coherence \cite{xxs,Ettefaghi,Ettefaghi2},
and quantum correlations\cite{alok16,ming20,bmd21}.


Among them, entanglement is the
most fundamental concept that can be rigorously quantified and characterized by the tools of QRTs. It has many potential applications in quantum information processing, including quantum cryptography \cite{horodecki,Ekert},
quantum teleportation \cite{Bennett93}, entanglement swapping \cite{Jennewein}, and so on.
There are many measures to quantify the entanglement in multiqubit quantum systems, such as the generalized geometric measure (GGM) \cite{Das, Sadhukhan}, three-$\pi$ entanglement measure \cite{Yong-Cheng Ou}, the genuinely multipartite concurrence (GMC) \cite{Ma},  and concurrence fill \cite{xiesongbo}. GGM is identified as an optimized distance of the
given state from the set of all states that are closest biseparable states. The three-$\pi$ entanglement measure consider all the bipartite residual entangle quantify three-qubit entanglement, based on the negativity.
GMC is an intuitive measure of multipartite entanglement, based on the
concurrence. Its lower bound can be obtained by the powerful detection criteria.
Concurrence fill is introduced as a genuine tripartite entanglement measure in terms of the area of the concurrence triangle. It represents a global entanglement, which fulfills the following properties:
non-negativity, monotonicity, discriminance, normalization, smoothness, and convexity \cite{YinfeiLi}. For the GHZ state, the concurrence fill can reach its maximum of 1, since the lengths of the three edges of  concurrence triangle are all equal to 1.
For the W state, its maximum is $F_{123}=8/9\approx0.89$. In particular, $F_{123}^{max}=0.89$ for three-flavor electron NOs considered in this work.

Here, we quantify the quantumness in three-flavor NOs by several tripartite entanglement measures, including GGM, three-$\pi$, GMC, and concurrence fill in QRTs, and compare their performances with respect to the ratio between the distance and neutrino energy $L/E$. The definitions of concurrence fill and three-$\pi$ take all bipartite
entanglements into account, and thus their change are always smooth. For the three-flavor electron and muon neutrino oscillations, concurrence fill is always larger
than  three-$\pi$, suggesting that concurrence fill contains more quantum resource. The definitions of GGM and GMC are given by the minimal entanglement measures of bipartite
entanglement, which lead to several nonanalytical sharp peaks. Moreover, we use the concurrence triangles for three-qubit systems to visualize the difference between GMC and concurrence fill.
It is shown that concurrence fill is always larger than or equal to the GMC in both sources of NOs.
The results show that concurrence fill is a genuine tripartite measure of tripartite entanglement in three-flavor NOs.

The paper is arranged as follows. In Sec. \ref{sec2}, we introduce three-flavor neutrino model. In Sec. \ref{sec3}, we briefly
review several tripartite entanglement measures, including GGM, three-$\pi$, GMC, and concurrence fill. In Secs. \ref{sec4} and \ref{sec5}, we compare the performances of the four
tripartite entanglement measures in the three-flavor electron
and muon antineutrino oscillations, respectively.
Finally, a summary is concluded in Sec. \ref{sec6}.

\section{The three-flavor neutrino model}
\label{sec2}

In the standard three-flavor neutrino oscillation model, a linear superpositions of the mass eigenstates $|\nu_1\rangle$, $|\nu_2\rangle$, $|\nu_3\rangle$, constitutes the three flavor states $|\nu_e\rangle$, $|\nu_\mu\rangle$, and $|\nu_\tau\rangle$:
\begin{align}
|\nu_{\alpha}\rangle=\sum_{k}U_{\alpha k}|\nu_{k} \rangle,
\label{nf1}
\end{align}
where $k=1,2,3$ and $\alpha=e,\mu,\tau$, which represents the neutrino flavor state. The related $U_{\alpha k}$ is a $3\times 3$ unitary matrix, is called as the Pontecorvo-Maki-Nakagawa-Sakata matrix \cite{Maki}. It can be characterized by three mixing angles and a charge coujugation and parity (CP) violating phase,
\begin{align}
U\!\!=\!\left(\!\!
           \begin{array}{ccc}
             c_{12}c_{13} & \! s_{12}c_{13} & \! s_{13}e^{-i\delta_{CP}} \\
             -\!s_{12}c_{23}\!-\!c_{12}s_{13}s_{23}e^{i\delta_{CP}} & \!c_{12}c_{23}\!-\!s_{12}s_{13}s_{23}e^{i\delta_{CP}} &\! c_{13}s_{23}\\
             s_{12}s_{23}\!-\!c_{12}s_{13}c_{23}e^{i\delta_{CP}} &\! -\!c_{12}s_{23}\!-\!s_{12}s_{13}c_{23}e^{i\delta_{CP}} &\! c_{13}c_{23} \\
           \end{array}
         \!\!\right),
         \label{PMNS}
\end{align}
where $c_{ij}\equiv \cos\theta_{ij}$ and $s_{ij}\equiv \sin\theta_{ij}$ $(i,j=1,2,3)$. The CP violating phase has not yet been observed, so we neglect it for simplicity.
The state $|\nu_{k}(t)\rangle$ is the mass eigenstate of the free Dirac Hamiltonian $H$ with an positive energy $E_k$, and its time evolution of the mass eigenstates $|\nu_{k}\rangle$ is expressed as
\begin{align}
|\nu_{k}(t)\rangle=e^{-iE_kt/\hbar}|\nu_{k}(0)\rangle.
\label{nf2}
\end{align}
Substituting the Eqs. (\ref{nf1}) and (\ref{PMNS}) into Eq. (\ref{nf2}), we can get the time evolution of flavor neutrino states $|\nu_{\alpha}(t)\rangle$ as
\begin{align}
|\nu_{\alpha}(t)\rangle=a_{\alpha e}(t)|\nu_{e}\rangle+a_{\alpha \mu}(t)|\nu_{\mu}\rangle+a_{\alpha \tau}(t)|\nu_{\tau}\rangle,
\label{timeevo}
\end{align}
where $a_{\alpha \beta}(t)= \sum_kU_{\alpha k}e^{-iE_kt/\hbar}U_{\beta k}^\ast$. Finally, the transition probability for detecting $\beta$ neutrino in the original $\alpha$ neutrino state is given by \cite{Garcia2},
\begin{eqnarray}
P_{\alpha\beta}&=&\delta_{\alpha\beta}-4\sum_{k>l}\textrm{Re}(U_{\alpha k}^\ast U_{\beta k}U_{\alpha l}U_{\beta l}^\ast)\sin^2\left(\Delta m^2_{kl}\frac{Lc^3}{4\hbar E}\right)\nonumber\\
&+&2\sum_{k>l}\textrm{Im}(U_{\alpha k}^\ast U_{\beta k}U_{\alpha l}U_{\beta l}^\ast)\sin\left(\Delta m^2_{kl}\frac{Lc^3}{2\hbar E}\right),
\label{tp}
\end{eqnarray}
where $\Delta m^2_{kl}=m^2_{k}-m^2_{l}$, $E$ is the energy of the neutrino in neutrino experiments,
and $L\thickapprox ct$ ($c$ is the speed of light in free space) is the distance between the source and the detector.

For convenience, we can write the oscillatory quantity $\sin^2\left(\Delta m^2_{kl}\frac{Lc^3}{4\hbar E}\right)$ that appears in Eq. (\ref{tp}) as ~\cite{Giunti},
\begin{align}  
\sin^2\left(\Delta m^2_{kl}\frac{Lc^3}{4\hbar E}\right)=\sin^2\left(1.27\Delta m^2_{kl}[eV^2]\frac{L[km]}{E[GeV]}\right).
\end{align}
Meantime, the oscillation parameters in normal ordering of the neutrino mass spectrum ($m_1 < m_2 < m_3$) are chosen as
\begin{eqnarray}
\Delta m_{21}^2 &=& 7.50 \times {10^{ - 5}}e{V^2},\nonumber\\
\Delta m_{31}^2& = &2.457 \times {10^{ - 3}}e{V^2},\nonumber\\
\Delta m_{32}^2 &=& 2.382 \times {10^{ - 3}}e{V^2},\nonumber\\
{\theta _{12}} &=& {33.48^ \circ },  {\theta _{23}} = {42.3^ \circ },  {\theta _{13}} = {8.50^ \circ }.
\end{eqnarray}
The occupation number in the neutrinos used here can be established in the following correspondence \cite{mbf09}
\begin{eqnarray}
\left| {{\nu _e }} \right\rangle  &\equiv {\left| 1 \right\rangle _e} \otimes {\left| 0 \right\rangle _\mu } \otimes {\left| 0 \right\rangle _\tau } &\equiv \left| {100} \right\rangle,\\
\left| {{\nu _\mu }} \right\rangle  &\equiv {\left| 0 \right\rangle _e} \otimes {\left| 1 \right\rangle _\mu } \otimes {\left| 0 \right\rangle _\tau } &\equiv \left| {010} \right\rangle,\\
\left| {\nu_\tau } \right\rangle  &\equiv {\left| 0 \right\rangle _e} \otimes {\left| 0 \right\rangle _\mu } \otimes {\left| 1 \right\rangle _\tau } &\equiv \left| {001} \right\rangle.
\end{eqnarray}
Thus, the flavor oscillations of neutrino can be seen as the time evolution of a tripartite quantum state. Then, from Eq. (\ref{timeevo}), we have
\begin{align}
{\left| {\psi _\alpha (t)} \right\rangle } = {a_{\alpha e}}(t)\left| {100} \right\rangle  + {a_{\alpha \mu }}(t)\left| {010} \right\rangle  + {a_{\alpha \tau }}(t)\left| {001} \right\rangle.
\end{align}
As a result, we can use the framework of quantum resource theory to study the flavor oscillations of neutrino. In the following, we focus on the performances of the tripartite entanglement measures, including GGM, three-$\pi$, GMC, and concurrence fill, and give a detailed comparison among them in this three-flavor neutrino systems.

\section{Measure of Tripartite Entanglement}
\label{sec3}
\subsection{GGM}
The GGM \cite{Das, Sadhukhan} of an $N$-partite pure state $\left| \psi_{N} \right\rangle$, can measure the entanglement of pure quantum
states of an arbitrary number of parties. It is defined as an optimized distance of the
given state from the set of all states that are not genuinely, given by
\begin{align}
\varepsilon(\left| \psi_{N} \right\rangle)=1- \Lambda^2_{max}(\left| \psi_{N} \right\rangle),
\label{Eq.GGM1}
\end{align}
where $\Lambda_{\max}(\left|\, \psi_{N} \,\right\rangle)=\max|\langle \chi\,|\,\psi_{N}\rangle|= \mathop {\max} \limits_{\left|\chi\right\rangle} F (\left|\psi_N\right\rangle, \left|\chi\right\rangle )$.
The maximization is performed over all $\left|\chi\right\rangle$ that are not multiparty entangled. $F(\left|\psi_N\right\rangle, \left|\chi\right\rangle)$ donates the fidelity between two pure states $\left|\psi_N\right\rangle$ and $\left|\chi\right\rangle$.
Based on the Hilbert-Schmidt distances, we can obtain an equivalent mathematical expression of Eq. (\ref{Eq.GGM1})
\begin{align}
G(\left| \psi _{n}\right\rangle)=1-max\{\lambda^2_{I:L}|I\cup L=\{A_1, ..., A_N\},I \cap L= \emptyset\},
\label{Eq.GGM2}
\end{align}
where $\lambda_{I:L}$ is the maximal Schmidt coefficient in the bipartite split $I:L$ of $\left| \psi_{N} \right\rangle$.

\subsection{The three-$\pi$ entanglement}
For a three-partite pure state $\left| \psi \right\rangle_{ABC}$, the Coffman-Kundu-Wootters-inequality-like \cite{Osborne}  monogamy inequality by the negativity quantified the entanglement is given by
\begin{align}
N^2_{AB}+N^2_{AC}\leq N^2_{A(BC)},
\label{Eq.threeentanglement1}
\end{align}
where $N_{AB}$ and $N_{AC}$ are the sum of the negative eigenvalues of the partial transpose of the states $\rho_{AB}=\textrm{Tr}_C(\rho_{ABC})$ and $\rho_{AC}=\textrm{Tr}_B(\rho_{ABC})$, respectively.
Following the idea of Ref \cite{Yong-Cheng Ou}, we have
\begin{align}
N_{A(BC)}= C_{A(BC)}=\sqrt {2[1-\textrm{Tr}(\rho^2_A)]},
\label{Eq.threeentanglement2}
\end{align}
where $\rho_A=\textrm{Tr}_{BC}(\rho_{ABC})$.
The residual entangle is defined as the difference between the two sides of Eq. (\ref{Eq.threeentanglement1})
\begin{align}
\pi_A=N^2_{A(BC)}-N^2_{AB}-N^2_{AC}.
\label{Eq.threeentanglement3}
\end{align}
Similarly, if one takes the different focus $B$ and $C$, the other two residual entangle are created
\begin{align}
\pi_B=N^2_{B(AC)}-N^2_{BA}-N^2_{BC},
\label{Eq.threeentanglement4}
\end{align}
\begin{align}
\pi_C=N^2_{C(AB)}-N^2_{CA}-N^2_{CB},
\label{Eq.threeentanglement5}
\end{align}
respectively. Note that the residual
entanglement for the different focus changes under transformations of the qubits. Finally, the three-$\pi$ entanglement of the tripartite systems are obtained as the average of $\pi_A$, $\pi_B$, $\pi_C$
\begin{align}
\pi_{ABC}=\frac{1}{3}(\pi_A+\pi_B+\pi_C).
\label{Eq.threeentanglement6}
\end{align}

\subsection{GMC}
The GMC is a computable measure quantifying the amount of multipartite entanglement based on the well-known concurrence \cite{Ma}.
For $n$-partite pure states $\left| \Psi \right\rangle\in H_1 \otimes H_2 \otimes \cdot\cdot\cdot \otimes H_n$, where $\rm {dim}(H_i)=d_i$, $i=1,2,...,n$, the GMC is defined as
\begin{align}
C_{GME}(\left| \Psi \right\rangle)= \mathop {\min }\limits_{{\gamma _i} \in \gamma }\sqrt{2[1-\textrm{Tr}(\rho^2_{A_{\gamma_i}})]}
\label{Eq.GMC1}
\end{align}
where $\gamma=\{\gamma_i\}$ represents the set of all possible bipartitions $\{A_i|B_i\}$  of  $\{1,2,...n.\}$. The GMC can also be generalized to the case of mixed states via the convex roof construction
\begin{align}
C_{GME}(\rho)=\mathop {\inf }\limits_{\{p_i, \left| \psi_i \right\rangle\} }\mathop {\sum }\limits_{i }p_iC_{GME}(\left| \psi_i \right\rangle),
\label{Eq.GMC2}
\end{align}
where the infimum is taken over all possible decompositions $\rho=\sum_i\left| \psi_i \right\rangle \left\langle \psi_i  \right|$.
For three-qubit systems, GMC is exactly the square root of the length of the shortest edge of the concurrence triangle. Following the idea of Ref. \cite{xiesongbo}, hereinafter we also ignore the square root and treat $C_{GME}$ as the length of the shortest edge for simplicity.

\subsection{The concurrence fill}
In 2021, Xie \emph{et al}. introduced the method of concurrence fill to genuinely capture triangle measure of tripartite entanglement, based on Heron¡¯s formula. Compared to other tripartite entanglement measures,  concurrences fill has the two advantages: (i) it contains more information; (ii) it is always smooth, while other measures contain a minimum argument, which will leads to a nonanalytical sharp peaks. In their paper, a concurrence triangle is presented that three squared one-to-other concurrences are the lengths of the three edges of a triangle for a three-qubit system. Then concurrence fill is defined as the square root of the area of the concurrence triangle. This gives the following expression of genuine tripartite entanglement measure \cite{xiesongbo}
\begin{align}
F_{123}\equiv\left[\frac {16}{3}Q(Q-C^2_{1(23)})(Q-C^2_{2(13)})(Q-C^2_{3(12)})\right]^{1/4},
\label{Eq.concurrence fill1}
\end{align}
where
\begin{align}
Q=\frac{1}{2}(C^2_{1(23)}+C^2_{2(13)}+C^2_{3(12)}),
\label{Eq.concurrence fill1}
\end{align}
with $C^2_{1(23)}$, $C^2_{2(13)}$, and $C^2_{3(12)}$ being $2[1-\textrm{Tr}(\rho^2_A)]$, $2[1-\textrm{Tr}(\rho^2_B)]$, and $2[1-\textrm{Tr}(\rho^2_C)]$, respectively.
$Q$ is the half-perimeter, the prefactor $16/3$ ensures the normalization $0\leq F_{123} \leq 1$, and the extra square root beyond
Heron¡¯s formula guarantees local monotonicity under local
quantum operations assisted with classical communications.

\section{Entanglement in electron antineutrino oscillations}
\label{sec4}
When the electron neutrino is generated at initial time $t=0$,
the evolutive states for three-flavor NOs is given by
\begin{align}
{\left| {\psi _e (t)} \right\rangle } = {a_{ee}}(t)\left| {100} \right\rangle  + {a_{e\mu }}(t)\left| {010} \right\rangle  + {a_{e\tau }}(t)\left| {001} \right\rangle.
\label{Eq.e1}
\end{align}

To quantify the tripartite entanglement, we shall focus on the the density matrix  $\rho _{ABC}^e(t)={\left| {\psi _e (t)} \right\rangle}\left\langle\psi _e (t)\right|$, that is,
\begin{eqnarray}
\rho _{ABC}^e(t) = \left( {\begin{array}{*{20}{c}}
0 & 0 & 0 & 0 & 0 & 0 & 0 & 0\\
0 & {\rho _{22}^e} & {\rho _{23}^e} & 0 & {\rho _{25}^e} & 0 & 0 & 0\\
0&{\rho _{32}^e}&{\rho _{33}^e}&0&{\rho _{35}^e}&0&0&0\\
0&0&0&0&0&0&0&0\\
0&{\rho _{52}^e}&{\rho _{53}^e}&0&{\rho _{55}^e}&0&0&0\\
0&0&0&0&0&0&0&0\\
0&0&0&0&0&0&0&0\\
0&0&0&0&0&0&0&0
\end{array}} \right),
\label{Eq.e2}
\end{eqnarray}
where the matrix elements are written as\\
$\rho _{22}^e\! =\!{\left| {{a_{e\tau }}(t)} \right|^2};$\,\,\,\;\;\;\;\;\,\;\,$\rho _{23}^e(t)\! =\! {a_{e\tau }}(t)a_{e\mu }^ * (t);$\,\;\;\,\,$\rho _{25}^e(t) \!= \!{a_{e\tau }}(t)a_{ee}^ * (t);$\\
$\rho _{32}^e \!= \!{a_{e\mu }}(t)a_{e\tau }^ * (t);$\,\,\,\,\;\,$\rho _{33}^e(t) \!= \!{\left| {{a_{e\mu }}(t)} \right|^2};$\,\;\;\;\;\;\;\,\,\,$\rho _{35}^e(t) \!= \!{a_{e\mu }}(t)a_{ee}^ * (t);$\\
$\rho _{52}^e \!=\! {a_{ee}}(t)a_{e\tau }^*(t);$\,\,\,\,\;\,$\rho _{53}^e(t) \!= \!{a_{ee}}(t)a_{e\mu }^*(t);$\,\,\;\;\;$\rho _{55}^e(t)\! = \!{\left| {{a_{ee}}(t)} \right|^2}.$

\begin{figure}[t]
\begin{center}
\includegraphics[width=7.8 cm,angle=0]{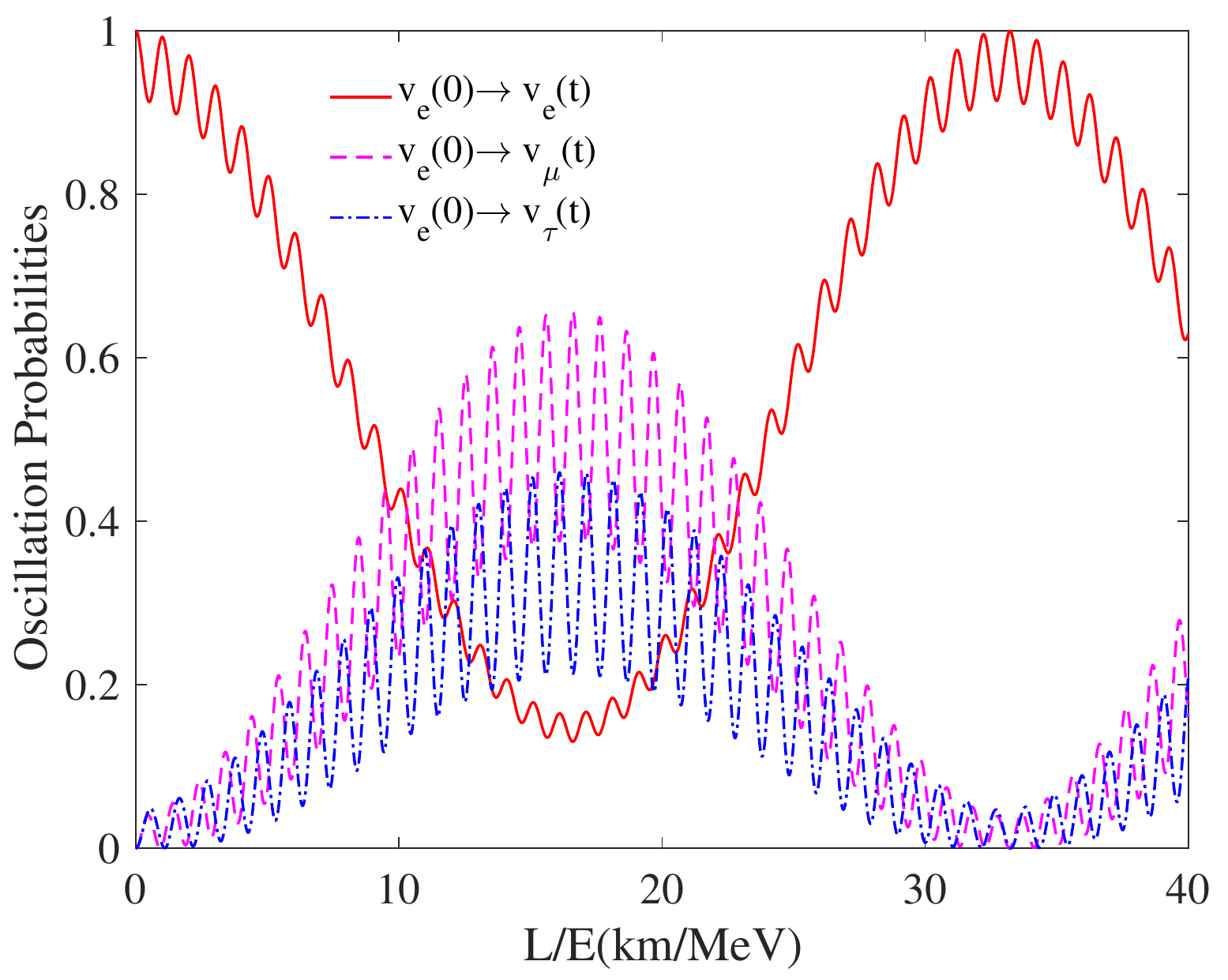}
\caption{Oscillation probabilities as a function of the ratio between the traveled distance $L$ and the energy $E$. The figure plots the oscillation probability ${v_e}(0) \to {v_e}(t)$ (red, solid line), ${v_e}(0) \to {v_\mu}(t)$ (purple, dashed line), ${v_e}(0) \to {v_\tau}(t)$ (blue, dashed-dotted line) when the initial neutrino flavor is electron flavor.}\label{fig1}
\end{center}
\end{figure}

And the corresponding oscillation probabilities, $P_{ee}(t)=|a_{ee}(t)|^2$, $P_{e\mu}(t)=|a_{e\mu}(t)|^2$ and $P_{e\tau}(t)=|a_{e\tau}(t)|^2$, in electron NOs as a function of $L/E$ are plotted in Fig. \ref{fig1}. When the initial neutrino is in electron flavor neutrino, the survive probability for maintaining in this flavor is always higher than $0.1$, and the other two transition probabilities
are smaller than $0.7$ in a range $[0, 40]$ of $L/E$ with dimension $\rm km/MeV$. The differences between survive and transition probabilities are large over a wide range with respect to $L/E$.

For the three-flavour neutrino state ${\left| {\psi _e (t)} \right\rangle }$ , using the Eq. (\ref{Eq.GGM2}), the GGM can be calculated as
\begin{align}
G(\rho^e)=1-max\{{\lambda^e _A},\, {\lambda^e _B},\, {\lambda^e _C}\},
\label{Eq.GGMe1}
\end{align}
where $\lambda^e_A$, $\lambda^e_B$ and $\lambda^e_C$ are the largest eigenvalues of the reduced density matrix $\rho^e_A$, $\rho^e_B$, $\rho^e_C$, respectively. It can found that $\lambda^e_A=max\{P_{e\tau}+P_{e\mu}, P_{ee} \}$, $\lambda^e_B=max\{P_{ee}+P_{e\tau}, P_{e\mu}\}$, $\lambda^e_C=max\{P_{ee}+P_{e\mu}, P_{e\tau} \}$, respectively.

To get the three-$\pi$ entanglement of the three-qubit electron neutrino system, we should firstly trace over the qubit $C$ and $B$ of $\rho_{ABC}$ to get the reduced density matrices: $\rho_{AB}$ and $\rho_{AC}$, respectively. The partial transpose $\rho_{AB}^{T}$ and $\rho_{AC}^{T}$ are obtained by focus on $A$. By solving the corresponding eigen equation, we can find the negativity of $N_{AB}$ and $N_{AC}$ with their negative eigenvalues. Using Eqs.(\ref{Eq.threeentanglement2}) and Eqs.(\ref{Eq.threeentanglement3}), we obtain the residual entangle $\pi_A$. In a similar way, we have $\pi_B$ and $\pi_C$. Finally, the systemic three-$\pi$ entanglement can be calculated as
\begin{align}
\pi_{ABC}(\rho^e)=\frac{4}{3}\Bigg[&-P^2_{ee}-P^2_{e\mu}-P^2_{e\tau}+P_{ee}\sqrt{P^2_{ee}+4P_{e\mu}P_{e\tau}}\nonumber \\
&+P_{e\mu}\sqrt{P^2_{e\mu}+4P_{ee}P_{e\tau}}+P_{e\tau}\sqrt{P^2_{e\tau}+4P_{ee}P_{e\mu}}\Bigg].
\label{pie1}
\end{align}

For GMC in the three-flavour electron neutrino system, from the Eq. (\ref{Eq.GMC1}), the GMC is expressed as $\textrm{min}\big\{2[1-\textrm{Tr}(\rho^2_A)], 2[1-\textrm{Tr}(\rho^2_B)], 2[1-\textrm{Tr}(\rho^2_C)]\big\}$. This gives
\begin{align}
C_{GME}(\rho^e)=\rm{min}\,\big\{&4P_{ee}\big(P_{e\mu}+P_{e\tau}\big)\,,\, 4P_{e\mu}\big(P_{ee}+P_{e\tau}\big)\,, \nonumber \\
&4P_{e\tau}\big(P_{ee}+P_{e\mu} \big)\big\}.
\label{CGMEe1}
\end{align}

\begin{figure}[t]
\begin{center}
\includegraphics[width=7.8 cm,angle=0]{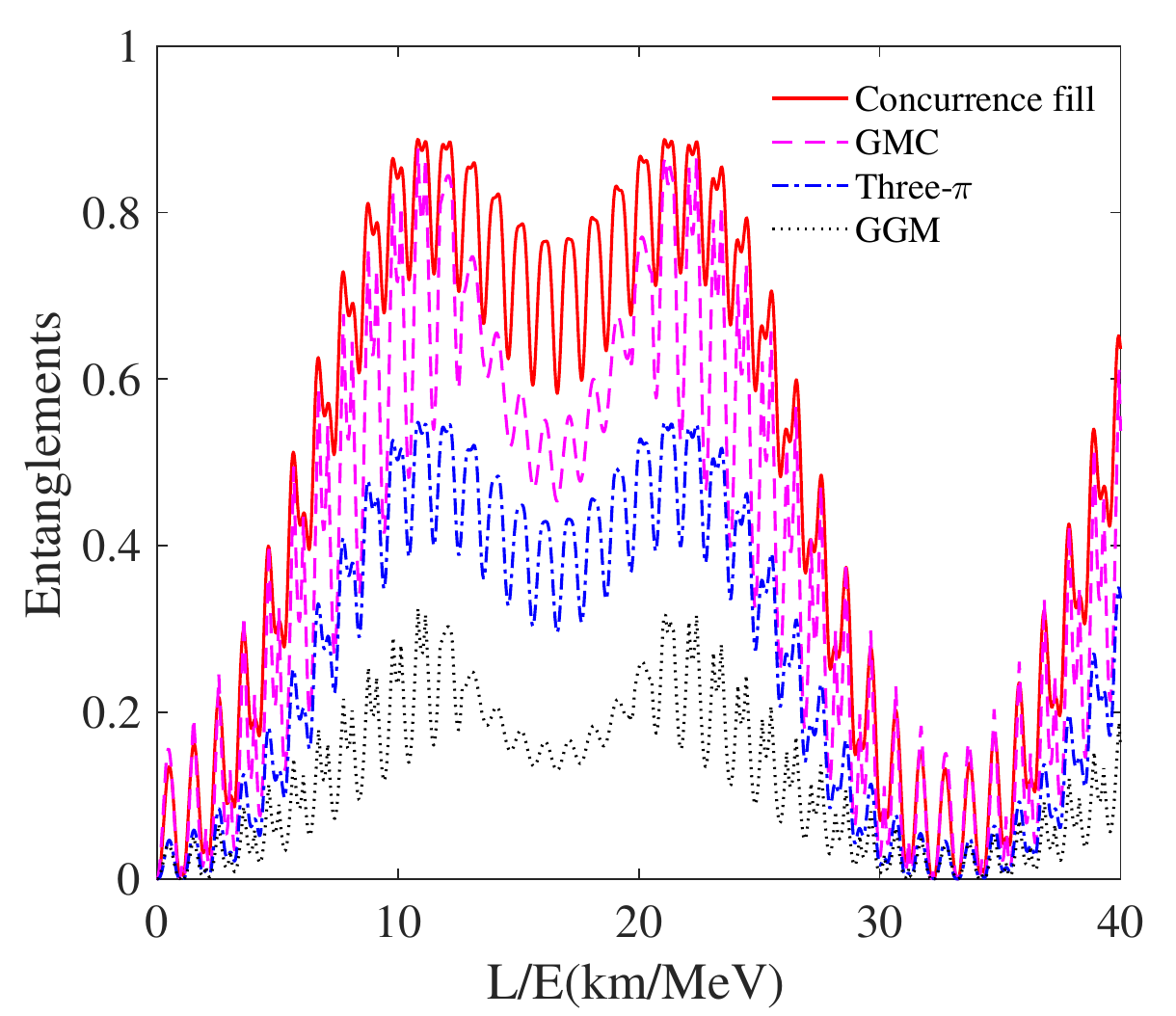}
\caption{The four kinds of multipartite entanglement measure, including GGM (black dotted line), three-$\pi$ (blue, dashed-dotted line), GMC (purple, dashed line), and concurrence fill (red, solid line) for the electron antineutrino oscillation system.}\label{fig2}
\end{center}
\end{figure}

Finally, by calculating the area of the concurrence triangle with the Eq. (\ref{Eq.concurrence fill1}), we can get concurrence fill
\begin{align}
F_{123}(\rho^e)=8\Bigg\{\frac{P^2_{ee}P^2_{e\mu}P^2_{e\tau}[P_{e\tau}P_{e\mu}+P_{ee}(P_{e\mu}+P_{e\tau})]}{3}\Bigg\}^{1/4}.
\label{Eq.concurrence fille1}
\end{align}
We can find that the amount of tripartite quantum entanglement depends on the three oscillation probabilities, which are subjected to the normalization constraint: $\sum_{\alpha}P_{\alpha\beta}=\sum_{\beta}P_{\alpha\beta}=1$ ($\alpha,\beta=e,\mu,\tau$). The maximal value of four tripartite entanglement measures: GGM, three-$\pi$, GMC, and concurrence fill, in three-flavored electron NOs is $1/3$, $4(\sqrt{5}-1)/9$, $8/9$, and $8/9$, respectively, when all the
oscillation probabilities are equal to $1/3$.

In Fig. \ref{fig2}, we plot the tripartite entanglement measures for three-flavored electron NOs as a function of ratio $L/E$, including GGM, three-$\pi$, GMC, and concurrence fill. At the point $L/E=0$, all the measure of multipartite entanglement are $0$. In particular, each edge of the concurrence triangle has a zero length, so the area of the concurrence triangle is also zero. Similarity, the four measures repeat a same change trend in one period of $L/E$ that increase firstly and then decrease. However, they take different amount of information in the neutrino propagation. Moreover,
the three oscillation probabilities get close to the critical condition that they take the same value, i.e., $P_{\alpha\beta}=1/3$, at around $L/E = 10.83~\rm km/MeV$. In this sense,
the maximal value of entanglement for these measures are approximatively $0.32$, $0.55$, $0.88$, and $0.89$, respectively. One can see that GMC and GGM have nonanalytical sharp peaks due to the nonanalytic minimum argument in its expression, concurrence fill is always smooth and contains more information. Although the three-$\pi$ presents a smooth curve, concurrence fill is always larger than it in the whole range of $L/E$. These results show that concurrence fill has advantages over other measures in the three-flavored electron NOs.

\begin{figure}[t]
\begin{center}
\includegraphics[width=9cm,angle=0]{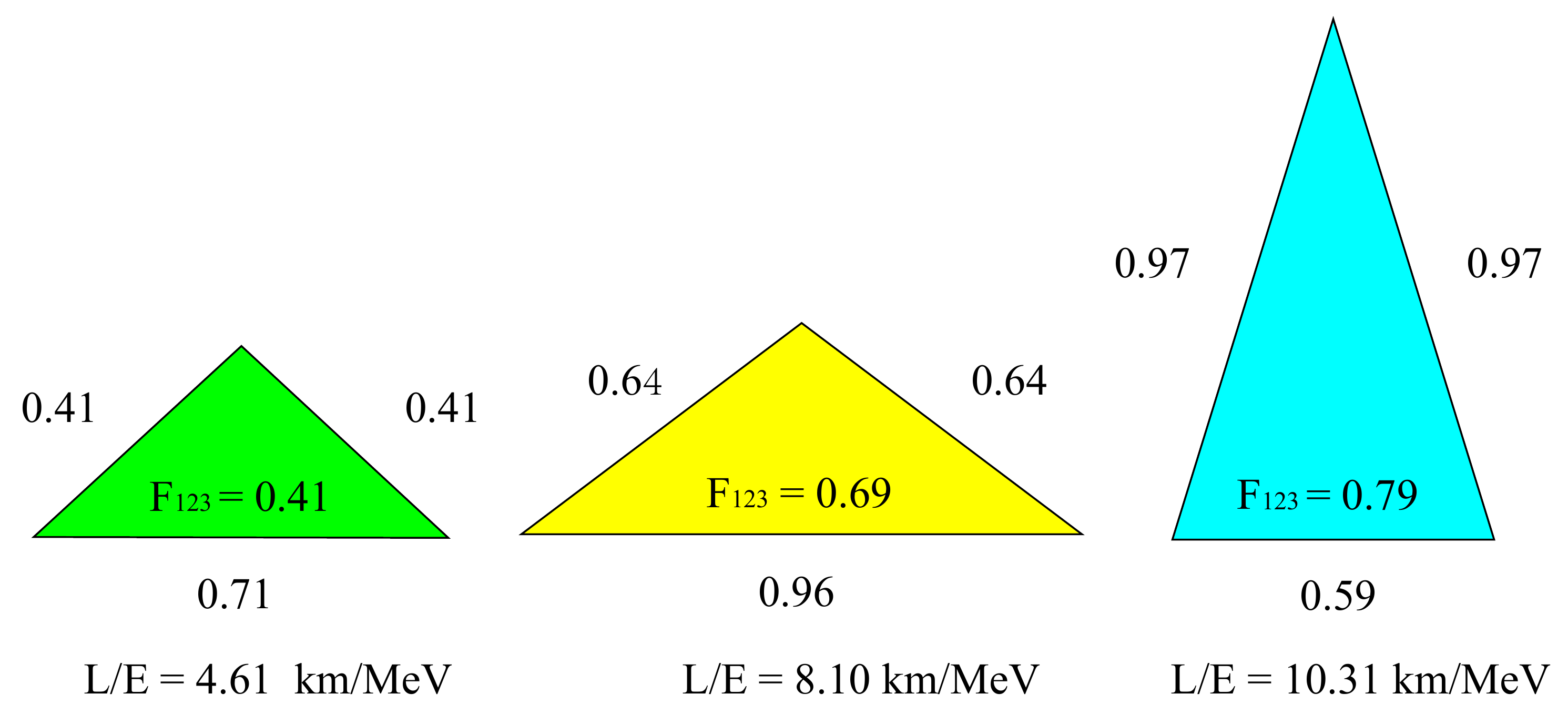}
\caption{The lengths of the edges and the areas for the concurrence triangles with $L/E = 4.61~\rm km/MeV$, $L/E = 8.10~\rm km/MeV$, and $L/E = 10.31~\rm km/MeV$, respectively, are shown in the three-flavored electron NOs.}\label{fige}
\end{center}
\end{figure}

In addition, to visualize the difference between GMC and concurrence fill for the electron NO systems, we make the use of the concurrence triangles to characterize the amount of information they carried.
The concurrence triangles for three cases: $P_{e\mu}=P_{e\tau}=0.115$, $P_{ee}=0.77$ at around $L/E = 4.61~\rm km/MeV$, $P_{e\mu}=P_{e\tau}=0.2$, $P_{ee}=0.6$ at around $L/E = 8.10~\rm km/MeV$, and $P_{ee}=P_{e\mu}=0.41$, $P_{e\tau}=0.18$ at around $L/E = 10.31~\rm km/MeV$, respectively, are plotted in Fig. \ref{fige}. They are all isosceles triangles, in which $F_{123}$ is the square root of the area of the concurrence triangle, and $C_{GME}$ is the corresponding shortest edge. One can see that concurrence fill is always larger than or equal to the GMC, meaning that concurrence fill possesses more information than GMC.

\section{Entanglement in muon antineutrino oscillations}

\label{sec5}
If the muon flavor state is prepared at initial time $t=0$,
the evolution of states for three-flavored NOs can be expressed as
\begin{align}
{\left| {\psi _\mu (t)} \right\rangle } = {a_{\mu e}}(t)\left| {100} \right\rangle  + {a_{\mu \mu }}(t)\left| {010} \right\rangle  + {a_{\mu\tau }}(t)\left| {001} \right\rangle.
\label{Eq.u1}
\end{align}

To quantify the entanglement measure, we shall focus on the the density matrix $\rho _{ABC}^\mu (t)={\left| {\psi_\mu (t)} \right\rangle }\left\langle\psi_\mu (t)\right|$. This gives
\begin{eqnarray}
\rho _{ABC}^\mu (t) = \left( {\begin{array}{*{20}{c}}
0&0&0&0&0&0&0&0\\
0&{\rho _{22}^\mu}&{\rho _{23}^\mu}&0&{\rho _{25}^\mu }&0&0&0\\
0&{\rho _{32}^\mu }&{\rho _{33}^\mu }&0&{\rho _{35}^\mu }&0&0&0\\
0&0&0&0&0&0&0&0\\
0&{\rho _{52}^\mu}&{\rho _{53}^\mu}&0&{\rho _{55}^\mu}&0&0&0\\
0&0&0&0&0&0&0&0\\
0&0&0&0&0&0&0&0\\
0&0&0&0&0&0&0&0
\end{array}} \right),
\label{Eq.u2}
\end{eqnarray}
where the matrix elements are given by\\
$\rho _{22}^\mu\!=\! {\left| {{a_{\mu \tau }}(t)} \right|^2}$; \; \;\;\;\;\,\,\,$\rho _{23}^\mu (t) \!=\! {a_{\mu \tau }}(t)a_{\mu \mu }^ * (t)$; \;\,\,$\rho _{25}^\mu(t)\!=\!{a_{\mu \tau }}(t)a_{\mu e}^ * (t);$\\
$\rho _{32}^\mu\!=\!{a_{\mu \mu }}(t)a_{\mu \tau }^ * (t)$; \;\,\,\,$\rho _{33}^\mu (t)\!=\!{\left| {{a_{\mu \mu }}(t)} \right|^2}$; \;\;\;\;\;\,\,\,$\rho _{35}^\mu (t)\!=\!{a_{\mu \mu }}(t)a_{\mu e}^ * (t);$\\
$\rho _{52}^\mu\!=\!{a_{\mu e}}(t)a_{\mu \tau }^*(t)$; \;\;\,\,$\rho _{53}^\mu (t)\! =\! {a_{\mu e}}(t)a_{\mu \mu }^*(t)$; \,\,\,$\rho _{55}^\mu (t)\!=\!{\left|{{a_{\mu e}}(t)}\right|^2}.$

\begin{figure}[t]
\begin{center}
\includegraphics[width=8 cm,angle=0]{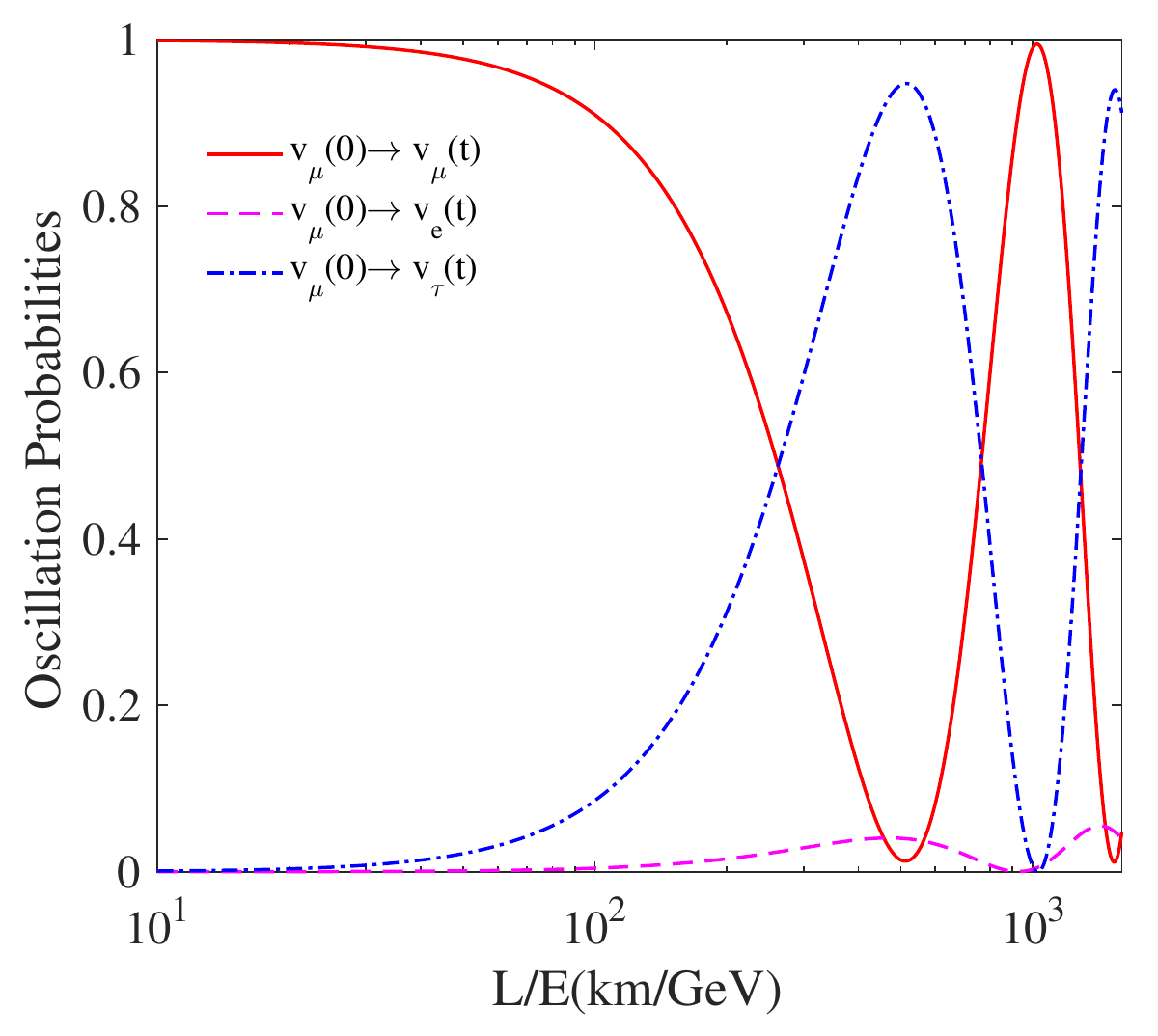}
\caption{Oscillation probabilities as a function of the ratio between the traveled distance $L$ and the energy $E$. The figure plots the oscillation probability ${v_\mu}(0) \to {v_\mu}(t)$ (red, solid line), ${v_\mu}(0) \to {v_e}(t)$ (purple, dashed line), ${v_\mu}(0) \to {v_\tau}(t)$ (blue, dashed-dotted line) when the initial neutrino flavor is muon flavor.}\label{fig3}
\end{center}
\end{figure}

The probabilities for finding the neutrino in state $|\nu_{e}\rangle$,
$|\nu_{\mu}\rangle$ and $|\nu_{\tau}\rangle$ are, respectively,
$P_{\mu e}(t)=|a_{\mu e}(t)|^2$, $P_{\mu\mu}(t)=|a_{\mu\mu}(t)|^2$ and $P_{\mu\tau}(t)=|a_{\mu\tau}(t)|^2$, for the initial muon flavor neutrino.
In Fig. \ref{fig3}, we show the oscillation probabilities of the muon neutrino oscillation as a function of $L/E$ with logarithmic scale. While the transition probability of $P_{\mu\tau}(t)$ shows
a relatively large range of variation, the transition probability of $P_{\mu e}(t)$ takes a trivial value with respect to $L/E$ with dimension $\rm km/GeV$.

For the three-flavour state ${\left| {\psi_ \mu (t)} \right\rangle }$ , the GGM is
\begin{align}
G(\rho^ \mu)=1-max\{\lambda^\mu_A,\, \lambda^\mu_B,\, \lambda^\mu_C\},
\label{GGMu2}
\end{align}
where $\lambda^\mu_A$, $\lambda^\mu_B$ and $\lambda^\mu_C$ are the largest eigenvalues of the reduced density matrix $\rho^\mu_A$, $\rho^\mu_B$, $\rho^\mu_C$, respectively. By calculation, $\lambda^\mu_A=max\{P_{\mu\tau}+P_{\mu\mu}, P_{\mu e} \}$, $\lambda^\mu_B=max\{P_{\mu e}+P_{\mu\tau}, P_{\mu\mu}\}$, $\lambda^ \mu_C=max\{P_{\mu e}+P_{\mu\mu}, P_{\mu \tau} \}$.

Similar to the case of electron neutrino oscillation system, we can calculate that the corresponding three-$\pi$ entanglement as
\begin{align}
\pi_{ABC}(\rho^ \mu)=\frac{4}{3}\Bigg[&-P^2_{\mu e}-P^2_{\mu \mu}-P^2_{\mu \tau}+P_{\mu e}\sqrt{P^2_{\mu e}+4P_{\mu \mu}P_{\mu \tau}}\nonumber \\
&+P_{\mu\mu}\sqrt{P^2_{\mu\mu}+4P_{\mu e}P_{\mu\tau}}+P_{\mu\tau}\sqrt{P^2_{\mu\tau}+4P_{\mu e}P_{\mu\mu}}\Bigg].
\label{piu1}
\end{align}
The GMC, as the smallest of three one-to-other bipartite entanglements of the three-flavour muon neutrino system, is expressed as
\begin{align}
C_{GME}(\rho^\mu)=\rm{min}\,\big\{&4P_{\mu e}\big(P_{\mu\mu}+P_{\mu\tau}\big)\,,\, 4P_{\mu\mu}\big(P_{\mu e}+P_{\mu\tau}\big) \nonumber \\
&4P_{\mu\tau}\big(P_{\mu e}+P_{\mu\mu} \big)\big\}.
\label{CGMEu1}
\end{align}
Concurrence fill, as a new method of measure the genuine multipartite entanglement, is obtained
\begin{align}
F_{123}(\rho^ \mu)=8\Bigg\{\frac{(P_{\mu e}P_{\mu\mu}P_{\mu\tau})^2[P_{\mu\tau}P_{\mu\mu}+P_{\mu e}(P_{\mu\mu}+P_{\mu\tau})]}{3}\Bigg\}^{1/4}.
\label{Eq.concurrence fillu1}
\end{align}

\begin{figure}[t]
\begin{center}
\includegraphics[width=8cm,angle=0]{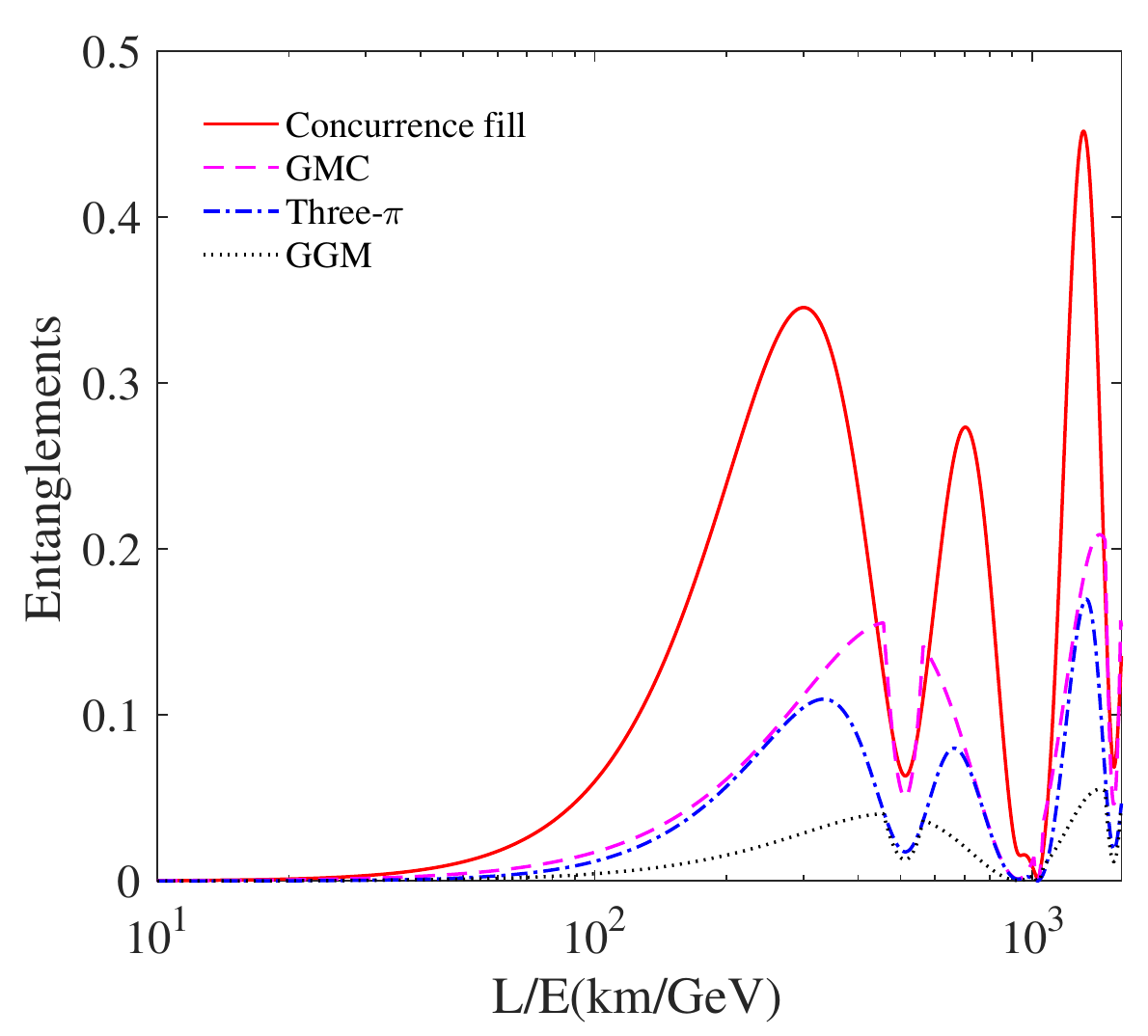}
\caption{In the muon antineutrino oscillation system, the multipartite entanglement measured by GGM (black dotted line), three-$\pi$ (blue, dashed-dotted line), GMC (purple, dashed line), and concurrence fill (red, solid line). }\label{fig4}
\end{center}
\end{figure}

\begin{figure}[t]
\begin{center}
\includegraphics[width=8cm,angle=0]{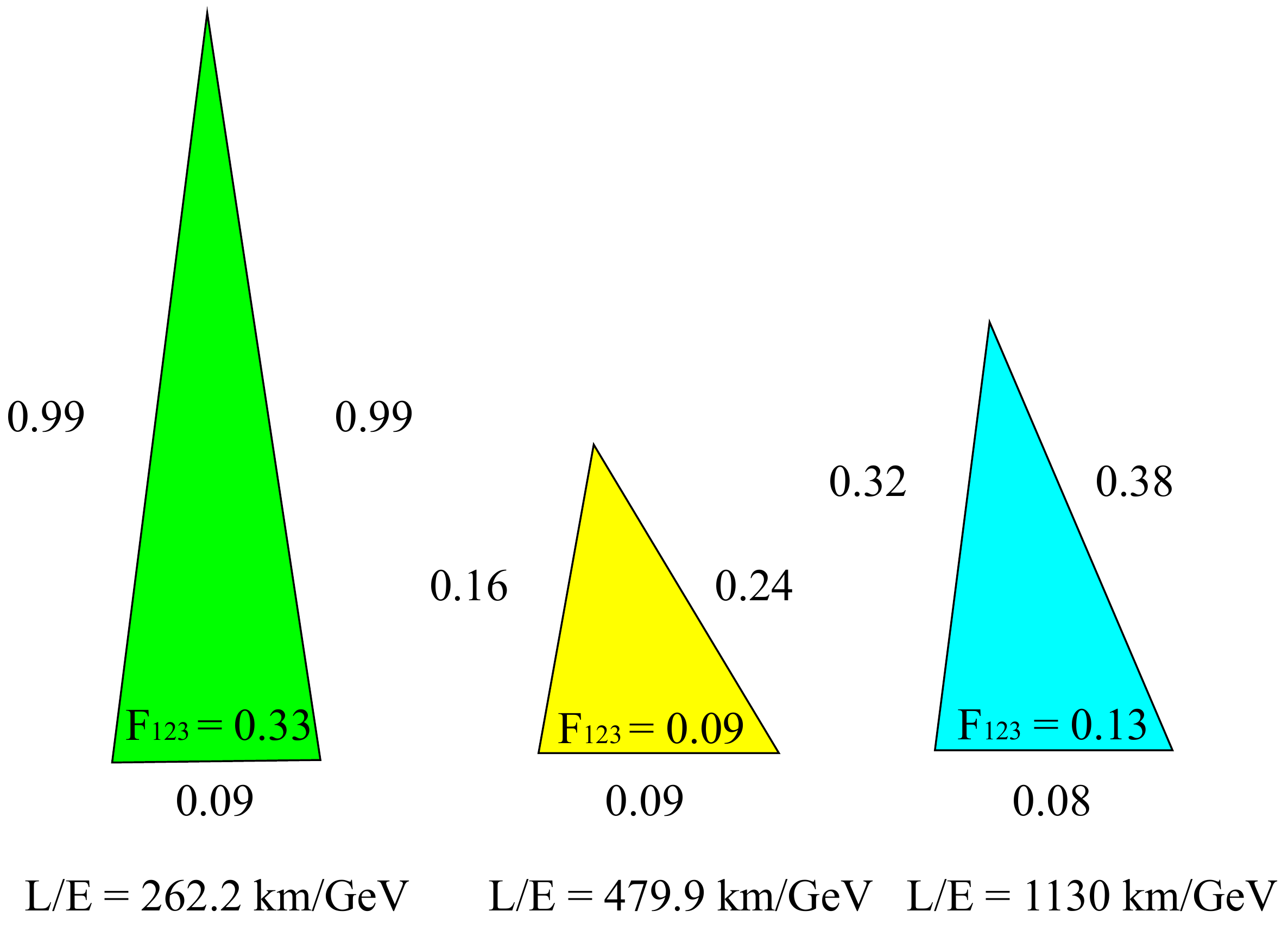}
\caption{The lengths of the edges and the areas for the concurrence triangles with $L/E = 262.2~\rm km/GeV$, $L/E = 479.9~\rm km/GeV$, and $L/E = 1130~\rm km/GeV$, respectively, are shown in the three-flavored muon NOs. }\label{figu}
\end{center}
\end{figure}

In Fig. \ref{fig4}, we plot GGM, three-$\pi$, GMC, and concurrence fill in three-flavored muon neutrino oscillations as a function of ratio $L/E$ with dimension $\rm km/GeV$. The minimum of the first concave interval of four entanglement measures are approximatively given by $0.13$, $0.18$, $0.51$, and $0.63$, respectively, when at around $L/E = 513.4~\rm km/GeV$ for three-flavored muon neutrino oscillations.
At the point $L/E=0$, only the survive probability is 1, and the other two transition probabilities are 0, and the four measures of multipartite entanglement are $0$. While GMC and GGM have six
nonanalytical sharp peaks, the change of concurrence fill are smooth in the range of $[10, 1600]$ of $L/E$. Also, the concurrence fill is always larger than GGM and three-$\pi$ in the whole change range of parameter.
Again, we can find that concurrence fill is a more genuine tripartite measure of tripartite entanglement possessing the advantage of smoothing change and more information simultaneously, compared to other three tripartite entanglement measures.

In Fig. \ref{figu}, we plot the concurrence triangles for the three-flavored muon neutrino systems in the case that $P_{\mu\mu}=P_{\mu\tau}=0.488$, $P_{\mu e}=0.024$ at around $L/E = 262.2~\rm km/GeV$ , $P_{\mu\mu}=0.022$, $P_{\mu\tau}=0.937$,
$P_{\mu e}=0.041$ at around $L/E = 479.9~\rm km/GeV$, and $P_{\mu\mu}=0.894, P_{\mu\tau}=0.087$, $P_{\mu e}=0.019$ at around $L/E = 1130~\rm km/GeV$, respectively. The $F_{123}$ and $C_{GME}$ are the square root of the area and shortest edge of the concurrence triangle, respectively. For the $L/E = 262.2~\rm km/GeV$, $L/E = 479.9~\rm km/GeV$, and $L/E = 1130~\rm km/GeV$, the square root of the area of the concurrence triangle are $0.33$, $0.09$, and $0.13$, respectively, and the corresponding shortest edges are $0.09$, $0.09$, and $0.08$. The shortest edges is always not more than the  square root of the area, suggesting that
concurrence fill contains more information than GMC.

\section{Summary}
\label{sec6}
In summary, we show that concurrence fill is a genuine tripartite measure to quantify the quantumness of three-flavor electron and muon neutrino oscillations. For initial electron-neutrino oscillations, concurrence fill can reach the maximum $0.89$, while the maximum of other three tripartite entanglement measures, including GGM, three-$\pi$, and GMC are $0.32$, $0.55$, and $0.88$, respectively.
It has been proven that concurrence fill is a more natural measure compared to other three measures. The
reasons are two-manifold: (i) Concurrence fill shows a smoothing change with respect to $L/E$, since it considers all the bipartite entanglement measures. GGM and GMC have nonanalytical sharp peaks due to the nonanalytic minimum argument in its expression, which lose sight of the global distribution of entanglement among the parties. (ii) Concurrence fill contains more quantum resource in the electron and muon neutrino propagation. Concurrence fill is always larger than or equal to the other three entanglement measures. All these features make concurrence fill a genuine information-theoretic quantification of tripartite quantum entanglement of quantum resource theories in three-flavor NOs.

\section*{ACKNOWLEDGMENTS}
This work is supported by the National Natural Science Foundation of China (Grants No. 12004006, No. 12075001, and No. 12175001), and Anhui Provincial Natural Science Foundation (Grant No. 2008085QA43).

\end{document}